\documentstyle[12pt,epsfig]{article}
\advance\textheight by \topskip
\begin{document}
\tolerance=100000
\thispagestyle{empty}
\setcounter{page}{0}

\newcommand{\be}{\begin{equation}}
\newcommand{\ee}{\end{equation}}
\newcommand{\br}{\begin{eqnarray}}
\newcommand{\er}{\end{eqnarray}}
\newcommand{\ba}{\begin{array}}
\newcommand{\ea}{\end{array}}
\newcommand{\bi}{\begin{itemize}}
\newcommand{\ei}{\end{itemize}}
\newcommand{\bn}{\begin{enumerate}}
\newcommand{\en}{\end{enumerate}}
\newcommand{\bc}{\begin{center}}
\newcommand{\ec}{\end{center}}
\newcommand{\ul}{\underline}
\newcommand{\ol}{\overline}
\newcommand{\ar}{\rightarrow}
\newcommand{\sm}{${\cal {SM}}$}
\newcommand{\mssm}{${\cal {MSSM}}$}
\newcommand{\susy}{{{SUSY}}}
\def\epem{\ifmmode{e^+ e^-} \else{$e^+ e^-$} \fi}
\newcommand{\Dir}{\kern -6.4pt\Big{/}}
\newcommand{\Dirin}{\kern -10.4pt\Big{/}\kern 4.4pt}
\newcommand{\DDir}{\kern -7.6pt\Big{/}}
\newcommand{\DGir}{\kern -6.0pt\Big{/}}
\newcommand{\eett}{$e^+e^-\rightarrow t\bar t $}
\newcommand{\eeZphi}{$e^+e^-\rightarrow Z\phi $}
\newcommand{\eeZH}{$e^+e^-\rightarrow ZH $}
\newcommand{\eeAH}{$e^+e^-\rightarrow AH $}
\newcommand{\eehww}{$e^+e^-\rightarrow hW^+W^-$}
\newcommand{\bbww}{$b\bar b W^+W^-$}
\newcommand{\eebbww}{$e^+e^-\rightarrow b\bar b W^+W^-$}
\newcommand{\ttbbww}{$t\bar t\rightarrow b\bar b W^+W^-$}
\newcommand{\eettbbww}{$e^+e^-\rightarrow t\bar t\rightarrow b\bar b W^+W^-$}
\newcommand{\bbb}{$ b\bar b$}
\newcommand{\ttb}{$ t\bar t$}
\def\Ord{\buildrel{\scriptscriptstyle <}\over{\scriptscriptstyle\sim}}
%
\def\OOrd{\buildrel{\scriptscriptstyle >}\over{\scriptscriptstyle\sim}}
\def\Ord{\buildrel{\scriptscriptstyle <}\over{\scriptscriptstyle\sim}}
\def\OOrd{\buildrel{\scriptscriptstyle >}\over{\scriptscriptstyle\sim}}
\def\pl #1 #2 #3 {{\it Phys.~Lett.} {\bf#1} (#2) #3}
\def\np #1 #2 #3 {{\it Nucl.~Phys.} {\bf#1} (#2) #3}
\def\zp #1 #2 #3 {{\it Z.~Phys.} {\bf#1} (#2) #3}
\def\pr #1 #2 #3 {{\it Phys.~Rev.} {\bf#1} (#2) #3}
\def\prep #1 #2 #3 {{\it Phys.~Rep.} {\bf#1} (#2) #3}
\def\prl #1 #2 #3 {{\it Phys.~Rev.~Lett.} {\bf#1} (#2) #3}
\def\mpl #1 #2 #3 {{\it Mod.~Phys.~Lett.} {\bf#1} (#2) #3}
\def\rmp #1 #2 #3 {{\it Rev. Mod. Phys.} {\bf#1} (#2) #3}
\def\sjnp #1 #2 #3 {{\it Sov. J. Nucl. Phys.} {\bf#1} (#2) #3}
\def\cpc #1 #2 #3 {{\it Comp. Phys. Comm.} {\bf#1} (#2) #3}
\def\xx #1 #2 #3 {{\bf#1}, (#2) #3}
\def\preprint{{\it preprint}}
\def\mathrm{\rm}

\begin{flushright}
{Cavendish-HEP-96/17}\\ 
{October 1996\hspace*{.5 truecm}}\\
\end{flushright}

\vspace*{\fill}

\begin{center}
{\Large 
$W^\pm + 4\mbox{jet}$ \bf{ 
irreducible background events from QCD\\
in top and Higgs searches
at the Next Linear Collider}}\\[1.5cm]
{\large Stefano Moretti}\\[0.4 cm]
{\it Cavendish Laboratory, University of Cambridge,}\\
{\it Madingley Road, Cambridge CB3 0HE, UK.}\\[0.25cm]
\end{center}
\vspace*{\fill}

\begin{abstract}
{\noindent 
\small 
Irreducible background effects due to $e^+e^-\ar
W^\pm + bbjj$ and $e^+e^-\ar W^\pm + jjjj$ events produced via QCD 
in top-antitop production and heavy
Higgs searches in the bremsstrahlung channel are studied
at the Next Linear Collider. Various  distributions
relevant to phenomenological analyses are given and compared to
those expected from top and Higgs signals in the decay channel
$b\bar bW^+W^-\ar b\bar b W^\pm jj$. This analysis follows similar
ones previously carried out for the case of the irreducible background
proceeding via electroweak interactions.}
\end{abstract}
\vskip1.0cm
\hrule
\vskip0.25cm
\noindent
Electronic mail: moretti@hep.phy.cam.ac.uk

\vspace*{\fill}
\newpage

\section*{1. Introduction}

In a series of recent papers \cite{pap1,pap2,pap3} various effects due to 
the irreducible background in \eebbww\ (with $W^+W^-\ar W^\pm+jj$) 
electroweak (EW)
events in top and Higgs searches
at the Next Linear Collider (NLC) were calculated, both 
within the Standard Model (\sm) \cite{pap1,pap3} and the Minimal Supersymmetric
Standard Model (\mssm) \cite{pap2}. 
On the one hand, top quarks are produced in pairs 
in electron-positron annihilations, through the mechanism
$e^+e^-\ar \gamma^*,Z^*\ar t\bar t$, and they subsequently decay
via $t\ar bW^\pm$ (if Supersymmetry is present also into, e.g., 
$t\ar bH^\pm$, via charged Higgses). On the other hand,
Higgs signatures\footnote{In the following
$\phi$ represents the \sm\ Higgs and $h,H$ and $A$ the \mssm\
(neutral) counterparts.} giving \bbww\ final states are
those produced in the processes \eeZphi\ (for the \sm, see 
Ref.~\cite{pap1}) and \eeZH, \eeAH, \eehww\ (for the \mssm, see
Ref.~\cite{pap2}), in which $Z$,  $A$ and $h$ decay into $b\bar b$ pairs, 
and $\phi, H\ar W^+W^-$. 

The relevance of non-resonant \eebbww\ EW diagrams, as well as their
interplay with the signals, in     the context of                       
top and Higgs searches at the NLC ($\sqrt s=350,500$ GeV), has been
demonstrated in the mentioned papers, to which we refer the reader
for details. In general,
irreducible background effects via EW interactions 
amount to several percents of the
total \eebbww\ cross section.
They are particularly visible in the spectrum of the $W^\pm$ momentum
at the $t\bar t$ threshold,
in the top-antitop excitation curve, as well as they can be relevant
for some of the distributions in the $b\bar b$ and $W^+W^-$
invariant masses. In all these cases, their knowledge is essential
in order to perform  at the NLC
the foreseen high precision measurements of top
and Higgs parameters \cite{ee500}.

It is the purpose of this short letter to calculate irreducible background
effects to the mentioned signals due to QCD events yielding the signature
$W^\pm +4\mathrm{jets}$ (in which the $W^\pm$ can decay either
hadronically or leptonically): that is, when the partons in the final state
are produced through the order $\alpha_s^2$, via gluon (G) interactions.
These can occur in 
\be\label{4q}
W^\pm+{q\bar q' Q\bar Q}\qquad\qquad\qquad\qquad\mathrm{(Fig.~1a)}
\ee  and 
\be\label{2q2g}
W^\pm+{q\bar q' G G}\qquad\qquad\qquad\qquad\mathrm{(Fig.~1b)}
\ee
partonic events. 
Since heavy flavour tagging (of $b$-quarks, essentially)
will constitute one of the most powerful experimental tools in hadronic 
phenomenology at the NLC\footnote{One should recall that 
the $b\bar b$ decay channel is largely dominant 
for Higgs bosons in the intermediate mass range, this rendering the detection
of such particles
rather problematic at the Large Hadron Collider (LHC) \cite{CMS,ATLAS},
because of the QCD background,
whereas the latter will not constitute a serious problem at the NLC.},
the signature of the above mechanisms is actually made up by the
following two components ($b$ indicates both a $b$ and a $\bar b$): 
\be\label{2b2j}
W^\pm +bb{jj},
\ee
and 
\be\label{4j}
W^\pm +{jjjj},
\ee
depending on the possible heavy (i.e., $b$) and light (i.e., $j$) quark
content of processes (\ref{4q})--(\ref{2q2g}).
Their relative importance in the total sample of $W^\pm +4\mathrm{jets}$
is dictated by the partonic production rates as well as
by the $b$-tagging efficiency/rejection  that
will be achieved by the NLC detectors.

Before exposing the plan of the paper, it is probably worth reminding
the reader the importance of the $W^\pm +4\mathrm{jet}$ signature 
arising from \bbww\ production and decay. In fact, two leptonic $W^\pm$ decays
would lead to a double disadvantage: first, a very much reduced statistics
and, second, problems in reconstructing invariant mass spectra because
of the two neutrinos escaping detection. Thus, at least one of the $W^\pm$'s
will most likely be tagged in the hadronic channel. Furthermore,
between the two possible
decays of the second $W^\pm$, it is usually preferred to resort
to the leptonic final state (i.e., $W^\pm\ar \ell\nu_\ell$, $\ell=e,\mu$),
more than to the hadronic  one  (i.e., $W^\pm\ar{jj}$). In fact, the 
experimental signature arising from the 
former has a few advantages with respect to the one from the latter:
it has a simpler topology inside the detectors
and thus it is easier to reconstruct;
it also allows one to get rid of complications due to the combinatorics
in case of a six jet final state; and finally, 
like the pure hadronic final state,
its kinematics is fully constrained (once the missing momentum
is assigned to the neutrino). 

We proceed in the rest of the paper as follows: the next Section is devoted 
to a brief description of our computational method and to the
declaration of the parameters used; the last one to
a discussion of the results with a brief summary in the end.

\section*{2. Calculation} 

The computation of the matrix elements of the elementary 
processes $e^+e^-\ar W^\pm+{q\bar q' Q\bar Q}$ (26 Feynman graphs) and 
$e^+e^-\ar W^\pm+{q\bar q' G G}$ (84 Feynman graphs), 
with $q^{(')},Q=u,d,s,c$ and $b$,
has been performed with the help of 
{\tt MadGraph} \cite{tim}, which uses the subroutines contained in 
{\tt HELAS} \cite{HELAS},
and the integrals over the phase spaces have been performed using
the package {\tt VEGAS} \cite{VEGAS}.

In the calculations 
we have adopted the following numerical values for the various parameters:
$$m_e=m_u=m_d=0,\qquad m_s=0.3~\mathrm{GeV},
\qquad m_c=1.4~\mathrm{GeV},
\qquad m_b=4.25~\mathrm{GeV},$$
$$M_Z=91.175~{\mathrm {GeV}},\quad\quad \Gamma_Z=2.5~{\mathrm {GeV}},$$
$$M_W=80.23~{\mathrm {GeV}},\quad\quad \Gamma_W=2.08~{\mathrm {GeV}},$$
\be\label{ewparam}
G_F=1.16637\times10^{-5}~{\mathrm {GeV}}^{-2},
\quad\quad\alpha_{em}\equiv \alpha_{em}(M_Z)= 1/128.
\ee
Note that the charged and neutral weak fermion--boson couplings are defined by
\be\label{GF}
g_W^2 = \frac{e^2}{\sin^2\theta_W} =  4 \sqrt{2} G_F M_W^{2}, \qquad
g_Z^2 = \frac{e^2}{\sin^2\theta_W\; \cos^2\theta_W} =  4 \sqrt{2} G_F M_Z^{2}.
\ee
For the vector and axial couplings
of the gauge bosons to the fermions, we use the `effective leptonic' value
\be\label{s2w_eff}
\sin^2(\theta_W)\equiv
\sin^2_{{eff}}(\theta_W)=0.2320.
\ee
The strong coupling constant $\alpha_s$  has been evaluated
at two loops, with $N_f=5$ and $\Lambda_{\overline{{MS}}}=190$ MeV,
at the scale $Q^2=s$, yielding $\alpha_s(M_Z^2)=0.115$.
Finally, the centre-of-mass (CM) energies considered for the NLC are
$\sqrt s=350$ GeV (assuming that $m_t=174$ GeV \cite{newtop}) 
and $\sqrt s=500$ GeV. 

\section*{3. Results and conclusions}

For simplicity, the $W^\pm$ boson entering in the final states
(\ref{4q})--(\ref{2q2g}) has been kept on-shell in our calculations. 
Furthermore, to
allow for the detection of the four jets we have imposed that these
are all sufficiently energetic and well separated, by imposing the 
cuts $E_{\mathrm{j}}>5$ GeV and $\cos\theta_{\mathrm{jj}}<0.95$, 
for all kind of jets ${\mathrm{j}}=j,b$ in the
final states (\ref{2b2j})--(\ref{4j}). Finally, as reference values 
for the $b$-tagging efficiency $\epsilon_b$ and non-$b$ rejection factor 
$R_{\not{b}}$ we adopt the 
`conservative' values 0.5 and 50, respectively \cite{CMS,ATLAS}.
In carrying out the discussion presented in this Section we closely follow
those of Refs.~\cite{pap1,pap2}, which are based on 
the top and Higgs detection
strategies exposed in Refs.~\cite{top} and \cite{Higgs}.

In detail, the selection procedures of top-antitop signals at the NLC 
usually considered are (see Ref.~\cite{Bagliesi}):
{\it i})   to perform a scan in $\sqrt s$;
{\it ii})  to study the $W^\pm$ momentum spectrum;
{\it iii}) to reconstruct the invariant mass of the three-jet system
$t\ar bW^\pm\ar bjj$.
The first two methods are normally used at threshold ($\sqrt s\approx 2m_t$),
whereas the last one has been considered for studies far above that
($\sqrt s\gg 2m_t$). 
Those for Higgs selection are in general:
{\it i})  the `missing mass' analysis (for two-body Higgs production);
{\it ii}) the `direct reconstruction' method 
(for all Higgs production mechanisms).
They are both described in Ref.~\cite{grosse}, and the procedure is
always to reconstruct the Higgs resonance: in the first case via
the spectrum in the recoil mass 
$M_{\mathrm{recoil}}^2=[(p_{e^+}+p_{e^-})-(p_{b}+p_{\bar b})]^2$
(since $Z,A\ar b\bar b$); in the second
case directly from the Higgs decay products (in $\phi,H
\ar W^{-}W^{+} \ar W^\pm jj$). 

In Fig.~2 we scan the energy range $2m_t-10~{\mathrm{GeV}}\Ord\sqrt s
\Ord2m_t+10~{\mathrm{GeV}}$ by computing the cross sections for producing
events of the type $W^\pm +bb{jj}$ and $W^\pm +{jjjj}$ via QCD.
As stressed in Ref.~\cite{grosse}, a cut in the minimum value of the
hadronic mass $M_{h}\equiv 
M_{\mathrm{jets}}$ is generally helpful in suppressing the reducible 
background, and we have implemented it here\footnote{In contrast, 
note that such a constraint
is not that effective in eliminating the irreducible EW background,
see discussions in Ref.~\cite{pap1,pap2}.} \cite{Bagliesi}.
For a top mass of 174 GeV, we adopt the requirement $M_h>200$ GeV \cite{pap2}.
From Fig.~2 one notices that the cross sections for the QCD processes
(\ref{2b2j})--(\ref{4j}) are rather large before the implementation
of the $M_h$ selection cut. In fact, one should remember that, 
for $\sqrt s\approx 2m_t$, the rates for top-antitop
events are a few hundred femtobarns, and
the ones for the irreducible EW background are of 10 fb or so, with
cuts in energy and cosine not yet implemented (see 
Ref.~\cite{pap2}). 
The need for the hadronic mass constraint is then clear, if one notices that, 
e.g., for 
$\sqrt s =350$ GeV, the cross sections after cuts for
events of the type $W^\pm +bbjj$ and $W^\pm +{jjjj}$ are 0.16 and 22.40 fb,
respectively, whereas the 
corresponding total rates were at the beginning (i.e., 
before cuts) 4.75 and 934.20 fb !
By considering that a further rejection factor $R_{\not{b}}^2\approx2500$ 
against light quark and gluon jets will multiply in the end the rates 
of $W^\pm +{jjjj}$ events and that an overall efficiency
to tag two $b$-quarks $\epsilon_b^2\approx0.25$ will multiply the 
irreducible background rates containing two $b$'s, 
one should conclude that the total
background in $W^\pm+\mathrm{4jet}$ events via QCD amounts to a few percent
of the EW one, around the threshold at $\sqrt s=350$ GeV (compare
to the rates given in Refs.~\cite{pap1,pap2}).

Fig.~3 shows the distribution in the momentum of the (tagged) $W^\pm$ boson
for events of the type (\ref{2b2j})--(\ref{4j}). For reference, we also
plot the same spectrum in case of top-antitop events decaying
into $W^\pm + bbjj$ (including finite $\Gamma_t$ effects, see 
Refs.~\cite{pap1,pap2}). The same cuts 
(in $E_{\mathrm{j}}$, $\cos\theta_{\mathrm{jj}}$
and $M_h$) have been applied to all processes (QCD background and 
$t\bar t$). Like in the case of integrated rates,
no significant impact should be expected from QCD events in this
case, being the main smearing effects due to finite top width
and irreducible EW background \cite{pap1,pap2}.

At $\sqrt s=350$ GeV the bremsstrahlung channel $e^+e^-\ar Z\phi$ 
is the dominant \sm\
Higgs production mechanism, and (as
stressed in Refs.~\cite{pap1,pap2}) the 
channel $Z\phi\ar (b\bar b)(W^+W^-)$ might well be one of best
ways to detect a heavy Higgs, thanks to the expected performances of the
vertex detectors in triggering the $Z$ boson \cite{ideal}.
Furthermore, the mode
$Z\ar b\bar b$ has a branching ratio (BR) 
about five times larger than that into $\mu^+\mu^-$
or $e^+e^-$ and it is equally free from backgrounds coming from 
$W^\pm$ decays.
In Fig.~4a we plot the differential distribution in the invariant mass 
of all the possible $W^\pm {jj}$ combinations for
the signatures (\ref{2b2j})--(\ref{4j}), since the Higgs resonance
will be searched for in the decay chain $\phi\ar W^+W^-\ar W^\pm jj$.
For reference, in the same figure we have superimposed the rates
for the Higgs signal (as an example, in the \sm). We assume a detector
resolution of 10 GeV, such that the rates in Fig.~4a for the \sm\ Higgs
process are nothing else than the total cross section for
$e^+e^-\ar Z\phi\ar b\bar b W^+W^-\ar bb W^\pm jj$ divided
by ten (with the cuts in $E_{\mathrm{j}}$ and 
$\cos\theta_{\mathrm{jj}}$ implemented).
Since the Higgs and $W^\pm + bbjj$ rates in Fig.~4a must be multiplied by 
$\epsilon_b^2$ whereas those for $W^\pm + jjjj$ must be divided by 
$R_{\not{b}}^2$, it is clear that for \sm\ Higgs masses up to 250 GeV
or so the number of background events should not modify the chances
of Higgs detection. In fact, e.g., for $M_\phi=160$ GeV, one gets 
(after $b$-tagging)
$18$ events for the \sm\ Higgs  
process and practically none
from $W^\pm + bbjj$ and $W^\pm + jjjj$; whereas, 
for $M_\phi=240$ GeV, the corresponding rates are 
$4.5$ and $0.5$ events 
(assuming $\int{\cal L}dt=10$ fb$^{-1}$ of integrated luminosity per annum).
The same can be said also for the case of 
below threshold decays $\phi\ar W^{\pm }W^{\mp *}\ar W^\pm jj$, 
for Higgs masses $M_\phi\Ord 160$ GeV, where the total
number of signal events
is indeed much larger than that of $W^\pm+\mathrm{4jet}$ events, by
more than one order of magnitude\footnote{Note that
in the range 140 GeV $\Ord M_\phi\Ord2M_W$ 
the off-shell two-boson decay has a branching ratio 
comparable or even larger than that in $b\bar b$ pairs, making its
exploitation important.
In fact, in case of $\phi\ar b\bar b$ decays one would have 
combinatorial problems, because of the presence of four $b$-quarks
in the final state. Mistags are instead absent in the case of the 
$W^{\pm *}W^{\mp}$ channel (at least for high $b$-tagging performances).}.
For \sm\ Higgs masses larger than 250 GeV, the possibilities of Higgs detection
appear to  diminish
significantly. On the one hand, phase space effects due to
the limited collider energy available strongly suppress the Higgs 
production rates.
On the other hand, both the $W^\pm +\mathrm{4jet}$ 
backgrounds become competitive
with the signal\footnote{Indeed, it should also be noticed that for
$M_\phi\OOrd250$ GeV one has that $\Gamma_\phi\OOrd4$ GeV, so that
the Higgs width starts becoming similar to
the invariant mass resolution and not all Higgs events are contained
in a 10 GeV bin centered around the resonance. To increase the signal 
rates, one should consider additional bins. Inevitably, this procedure
enhances the relative number of background events.
This also means that the dotted curve slightly overestimates
the rates for $M_\phi\OOrd250$ GeV. 
Note that similar comments will hold also at $\sqrt s=500$ GeV, see Fig.~6a
later on.}.
Therefore, for $M_\phi\OOrd250$ GeV, not only would one need to exploit a higher
luminosity
option, but also would have less control on the backgrounds.
For example (assuming now that $\int{\cal L}dt=100$ fb$^{-1}$),
for a Higgs mass of 270 GeV, one gets the following number of
events: 1.7 for the signal, and about 7 for the total 
$W^\pm+\mathrm{4jet}$ background. 

In this context, however, one could always resort to a cut in the invariant
mass of the $b\bar b$ pair, as for the signal the corresponding spectrum
will peak at the $Z$ mass, whereas for the background the di-jet mass
distributions follow the behaviours displayed in Fig.~4b.
A cut around $M_Z$, for example $|M_{\mathrm{jj}}-M_Z|<15$ GeV (as advocated in
Ref.~\cite{grosse}), should be enough to remove the irreducible QCD backgrounds
in $W^\pm +\mathrm{4jet}$ events. 

For a higher energy NLC (i.e., $\sqrt s=500$ GeV), irreducible
background effects due to QCD events have no importance in top
searches, as can be clearly seen from Fig.~5. After $b$-tagging
suppression, the $W^\pm+\mathrm{4jet}$ rates
are more than four orders of magnitude 
below the top-antitop ones (for mass resolution of 10 GeV or so),
after the application of the constraint
$0.95\le x_E\le1.05$,
where $x_E=E_{bW\ar3{\mathrm{jets}}}/E_{\mathrm{beam}}$ \cite{grosse}.

In the case of Higgs searches at $\sqrt s=500$, the prospects
look even more promising than at lower energy. 
In fact, the number of 
$W^\pm + 4\mathrm{jet}$ background events via QCD is always much smaller than
that of the signal, for Higgs masses up to about
400 GeV (after $b$-tagging): that is, the kinematic border 
$\sqrt s-M_Z$ for Higgs production via the bremsstrahlung mechanism, 
see Fig.~6a. In addition, the usual cut in di-jet 
invariant mass to enhance the $Z\ar b\bar b$ decay of the $Z\phi$ signal
will reduce $W^\pm+\mathrm{4jet}$ rates to negligible levels (see Fig.~6b).
In fact, the invariant masses produced by gluon splitting tend to concentrate
towards low mass 
values (especially in $W^\pm + bbjj$ events, see graphs 1,2 \& 4 in Fig.~1a), 
being instead the Higgs rates dictated by a 2.5 GeV wide
Breit-Wigner distribution centered at $M_Z$.

In conclusion, effects due to
irreducible background events via QCD interactions
in processes of the type $e^+e^-\ar W^\pm + bbjj$ and 
 $e^+e^-\ar W^\pm + jjjj$, which produce signatures similar to those of
top and \sm\ heavy Higgs in the channels $e^+e^-\ar t\bar t\ar b\bar b
W^+W^-$ and $e^+e^-\ar Z\phi\ar b\bar bW^+W^-$ (with $W^+W^-\ar 
W^\pm jj$) at the NLC, are generally much smaller compared to those
due to $e^+e^-\ar b\bar bW^+W^-$ background
events proceeding via EW interactions.
This is largely due to the performances expected from
 the $b$-tagging devices
in suppressing the QCD background in light quark and gluon jets and to
the event selection procedures exploiting the resonant kinematics of the signal 
events. In fact, on the one hand, the  $W^\pm + jjjj$ 
production cross section via QCD is
much larger than that for $e^+e^-\ar b\bar b W^+W^-\ar W^\pm bbjj$ EW events,
on the other hand, the detected cross section (i.e., after flavour tagging
but before the kinematic selection)
of $W^\pm +4\mathrm{jet}$ events can be smaller by only one
order of magnitude respect to that of $t\bar t$ production, and
bigger than that of $Z\phi$ production.
However, in the very end, the only EW rates 
(as already computed
in Refs.~\cite{pap1,pap2}) should give reliable account of the 
effects due to the irreducible background in both top \cite{pap1}
and Higgs \cite{pap2} searches, provided that suitable kinematic
selection criteria of the signals are adopted.
To our opinion, it was nonetheless important to assess the
quantitative importance of QCD background events, because of the many and high
precision measurements of both top and Higgs parameters foreseen at
the linear colliders of the next generation. In particular,
this has been done here with an exact calculation of the relevant tree-level
matrix elements in perturbative QCD.
(We do not expect in fact that soft QCD phenomena in
the hadronisation processes could modify our main results.)
Furthermore, as reference example, we have concentrated here on the
case of the \sm\ Higgs boson \cite{pap1} only: however, our arguments 
are also valid in the \mssm\ \cite{pap2}. Even in the case
of more detailed studies of the (colour) structure of $b\bar bW^+W^-\ar
b\bar b \ell\nu_\ell jj$ events (with $\ell=e,\mu$), such as those
carried out in Ref.~\cite{pap3}, the effects due to the QCD irreducible
background are negligible compared to those due to the EW one.  
Finally, we have assumed throughout the analysis a simplified $b$-tagging
procedure, which did take into account neither the different probabilities in 
misidentifying, on the one hand, light quarks and gluons and, on the other
hand, $c$-quarks, as bottom quarks, nor 
the consequent combinatorics entering in the effective 
tagging efficiency. 
Anyhow, more sophisticated and realistic 
algorithms will certainly not change the conclusions drawn from this study.

\subsection*{Acknowledgements}

This work is supported in part by the
Ministero dell' Universit\`a e della Ricerca Scientifica, the UK PPARC,
and   the EC Programme
``Human Capital and Mobility'', Network ``Physics at High Energy
Colliders'', contract CHRX-CT93-0357 (DG 12 COMA).

\goodbreak

\section*{Figure Captions}

\begin{itemize}

\item[{[1]}] Relevant Feynman diagrams contributing at lowest
order to processes (1) and (2): 
a) $W^\pm+{q\bar q' Q\bar Q}$ case;
b) $W^\pm+{q\bar q' G G}$ case. Permutations of real and virtual
lines along the fermion lines are not shown. An internal wavy line
represents a $W^\pm$, a $\gamma$ and a $Z$, as appropriate.

\item[{[2]}] Cross section around the top-antitop threshold
$2m_t\approx 350$ GeV for the final states (3) and (4).
The underlying cuts 
$E_{\mathrm{j}}>5$ GeV and 
$\cos\theta_{\mathrm{jj}}<0.95$ have been implemented on all
(gluon, light and heavy quark) jets in the final states.
Upper curves are before cuts, lower curves are after the cut
$M_{h}>200$ GeV.

\item[{[3]}] Differential distribution in the momentum of the $W^\pm$ boson,
for events of the type (3) and (4), at $\sqrt s=350$ GeV.
The underlying cuts 
$E_{\mathrm{j}}>5$ GeV and $\cos\theta_{\mathrm{jj}}<0.95$ 
have been implemented on all
(gluon, light and heavy quark) jets in the final states.
The top selection
requirement $M_{h}>200$ GeV has been also implemented. 
The shaded histogram represents the rates obtained from the process
$e^+e^-\ar t\bar t\ar b\bar b W^+W^-\ar W^\pm +bb jj$ (for
$\Gamma_t\ne0$, see Refs.~\cite{pap1,pap2}), with $m_t=174$ GeV,
for the same choice of cuts in $E_{\mathrm{j}}$, 
$\cos\theta_{\mathrm{jj}}$ and $M_{h}$.

\item[{[4]}] Differential distribution in the invariant mass 
of all the possible: a) $W^\pm+\mathrm{2jet}$ combinations, b) 
2jet combinations,
for events of the type (3) and (4), at $\sqrt s=350$ GeV.
The underlying cuts 
$E_{\mathrm{j}}>5$ GeV and $\cos\theta_{\mathrm{jj}}<0.95$ 
have been implemented on all
(gluon, light and heavy quark) jets in the final states.
Bins are 10 GeV wide. The dotted line in a) represents the corresponding
spectra in the case of \sm\
Higgs events in the $Z\phi\ar b\bar b W^+W^-\ar bb W^\pm jj$ channel,
assuming a 10 GeV resolution on $M_\phi$ (note the onset of the $H\ar ZZ$
decay channel at $M_\phi\approx2M_Z$).

\item[{[5]}] Differential distribution in the invariant mass 
of all the possible $3\mathrm{jet}$ combinations, 
for events of the type (3) and (4), at $\sqrt s=500$ GeV.
The underlying cuts 
$E_{\mathrm{j}}>5$ GeV and $\cos\theta_{\mathrm{jj}}<0.95$ 
have been implemented on all
(gluon, light and heavy quark) jets in the final states.
The top selection
requirement $0.95<x_E<1.05$ has been also implemented. 
The shaded histograms represent the rates obtained from the process
$e^+e^-\ar t\bar t\ar b\bar b W^+W^-\ar W^\pm \mathrm{4jets}$ (for
$\Gamma_t\ne0$, see Refs.~\cite{pap1,pap2}), with $m_t=174$ GeV,
for the same choice of cuts in $E_{\mathrm{j}}$, 
$\cos\theta_{\mathrm{jj}}$ and $x_E$.
Black shadowing: `right' $W^\pm b$ combination;
dotted shadowing: `wrong' $W^\pm b$ combination (see Refs.~\cite{pap1,pap2}).

\item[{[6]}] Same as Fig.~4, at $\sqrt s=500$ GeV.

\end{itemize}
\vfill
\clearpage
\begin{figure}[p]
~\epsfig{{file=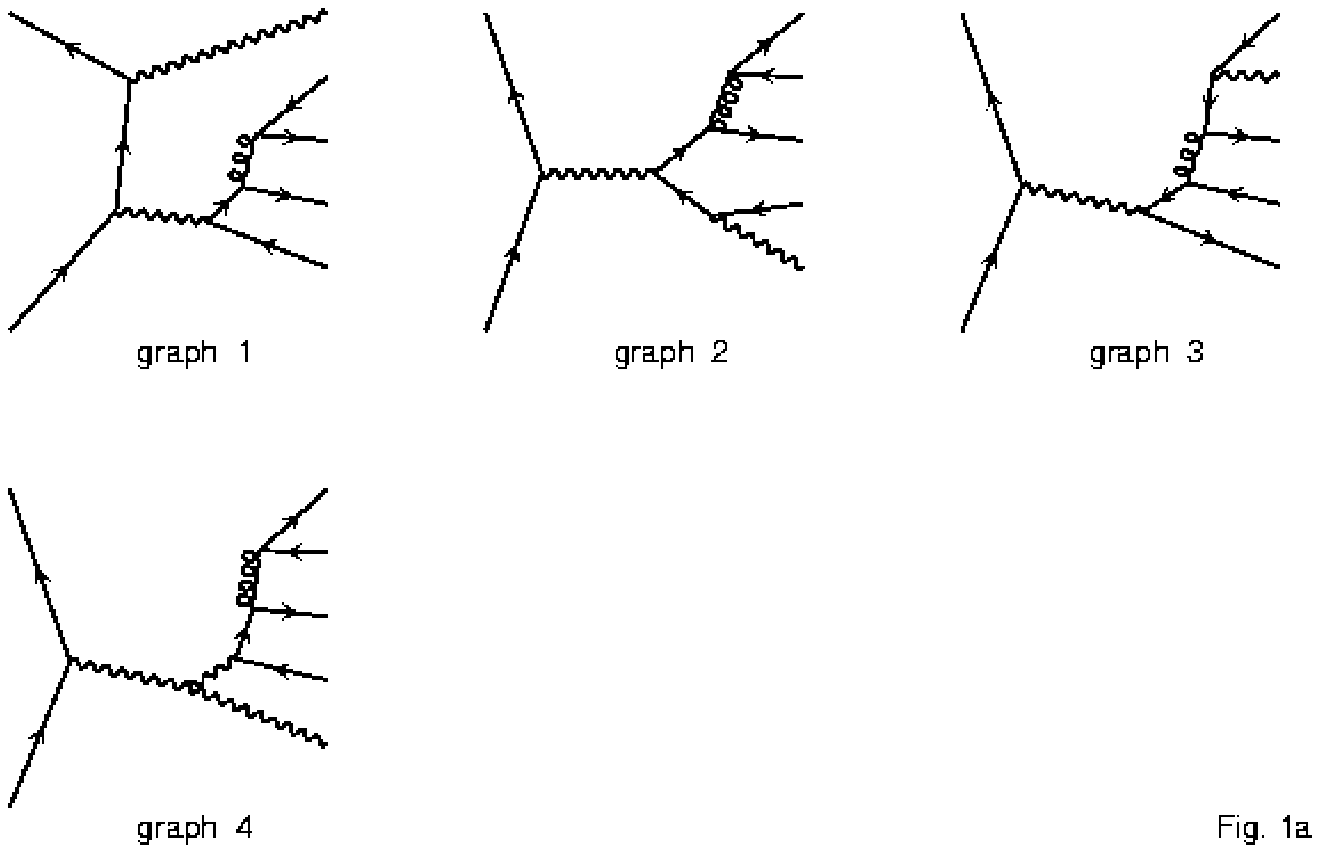,height=22cm}}  
\vspace*{2cm}
\end{figure}
\vfill
\clearpage

\begin{figure}[p]
~\epsfig{{file=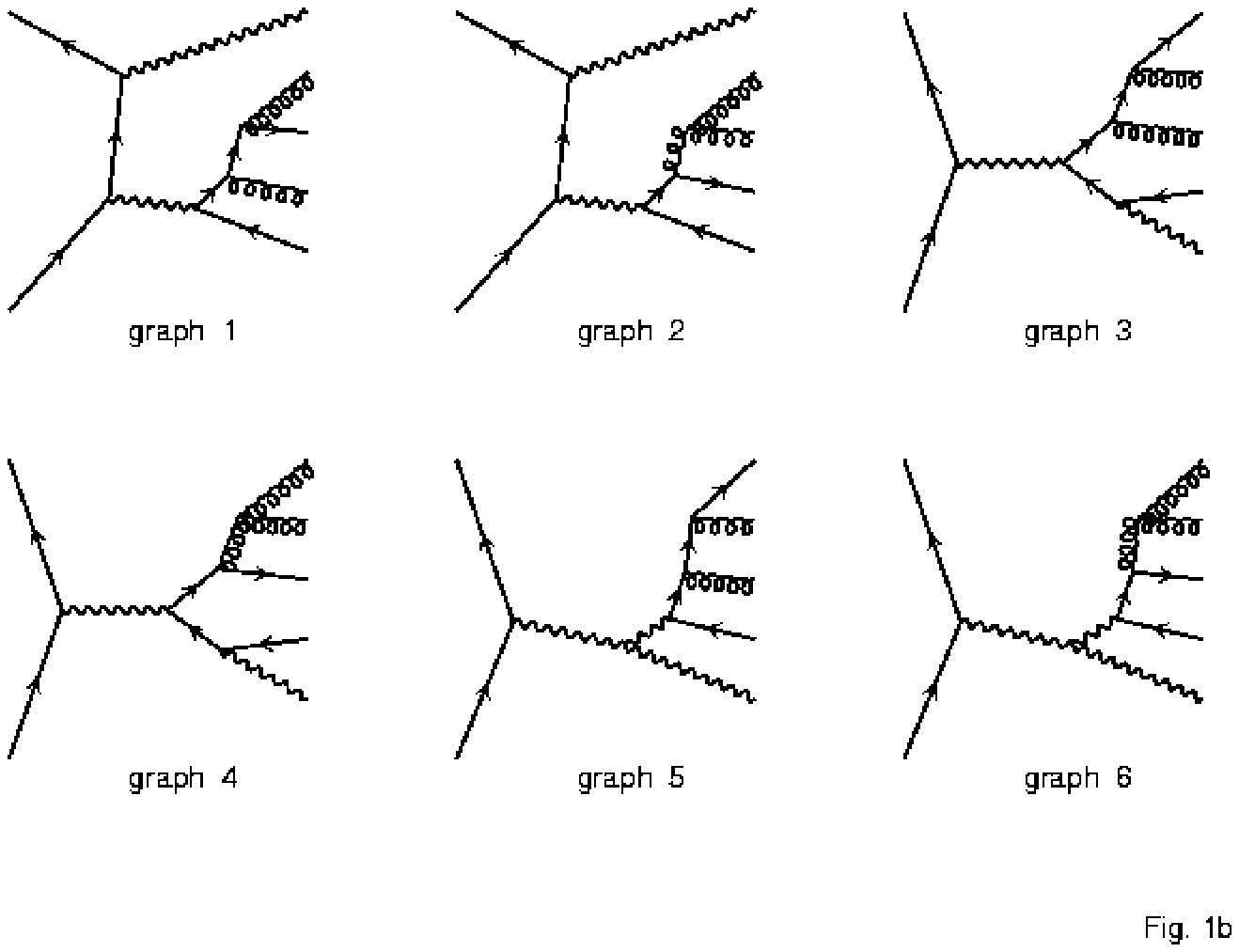,height=22cm}}  
\vspace*{2cm}
\end{figure}
\vfill
\clearpage
\begin{figure}[p]
~\epsfig{{file=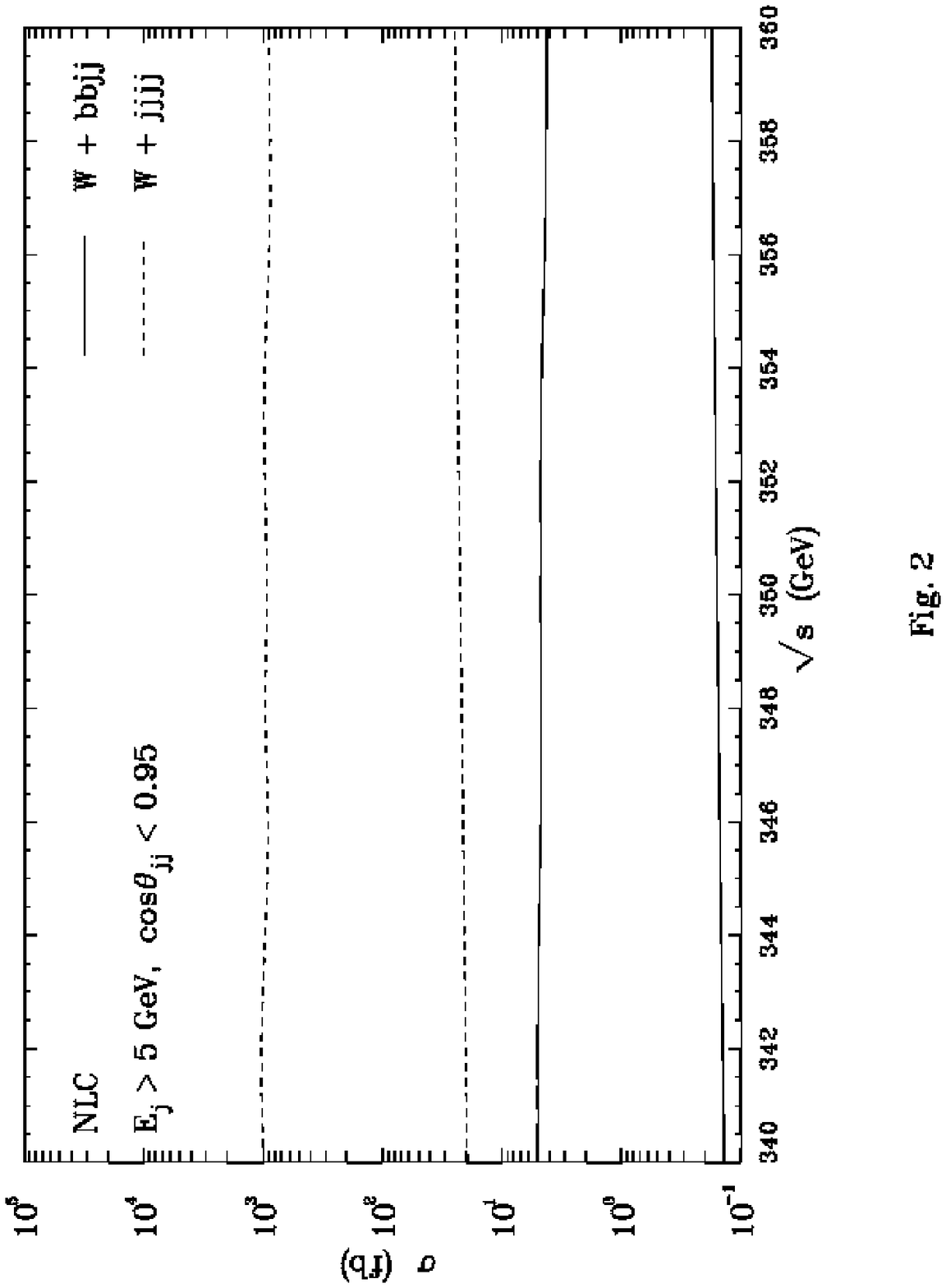,height=22cm}}  
\vspace*{2cm}
\end{figure}
\vfill
\clearpage
\begin{figure}[p]
~\epsfig{{file=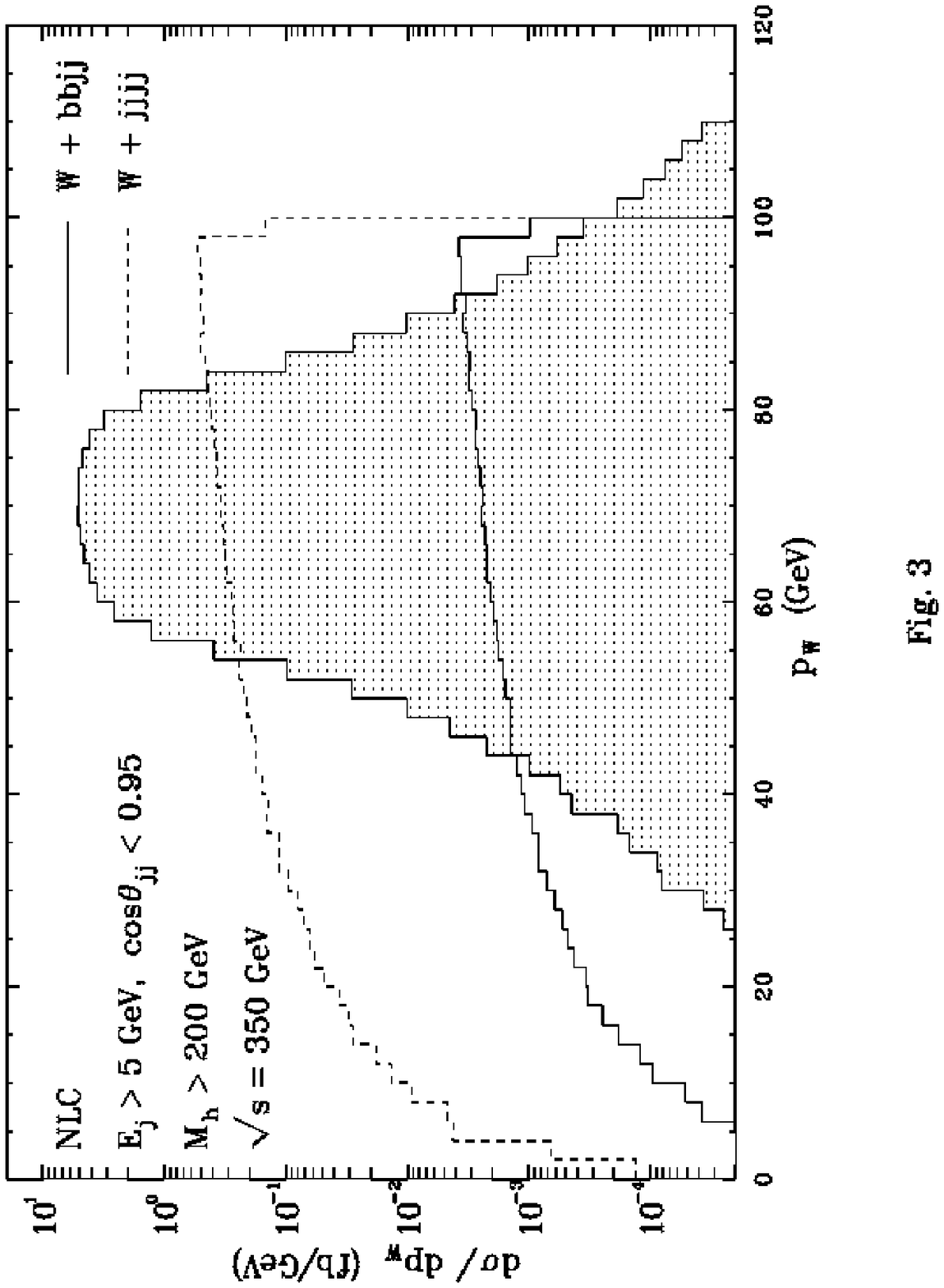,height=22cm}}  
\vspace*{2cm}
\end{figure}
\vfill
\clearpage
\begin{figure}[p]
~\epsfig{{file=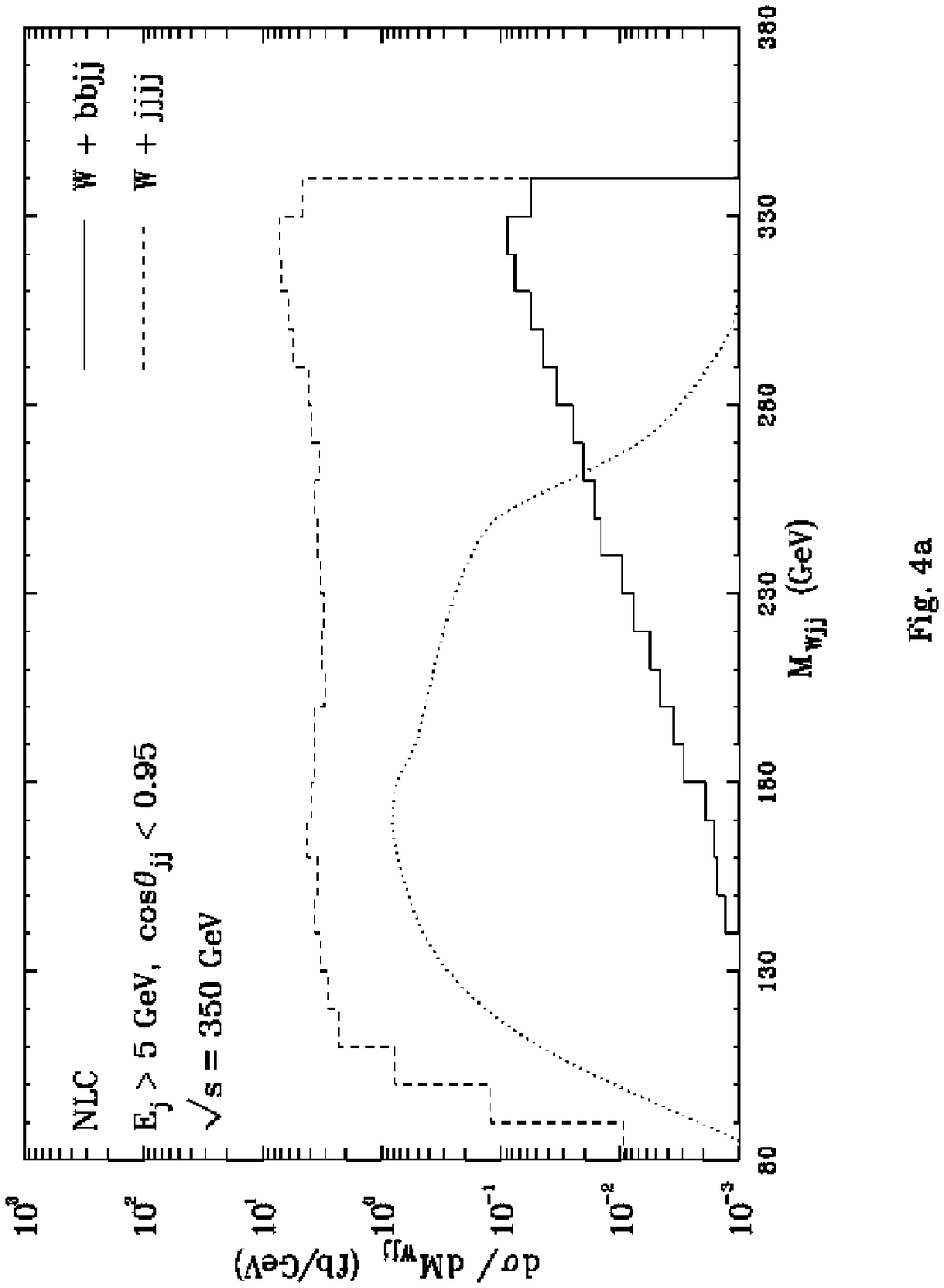,height=22cm}}  
\vspace*{2cm}
\end{figure}
\vfill
\clearpage
\begin{figure}[p]
~\epsfig{{file=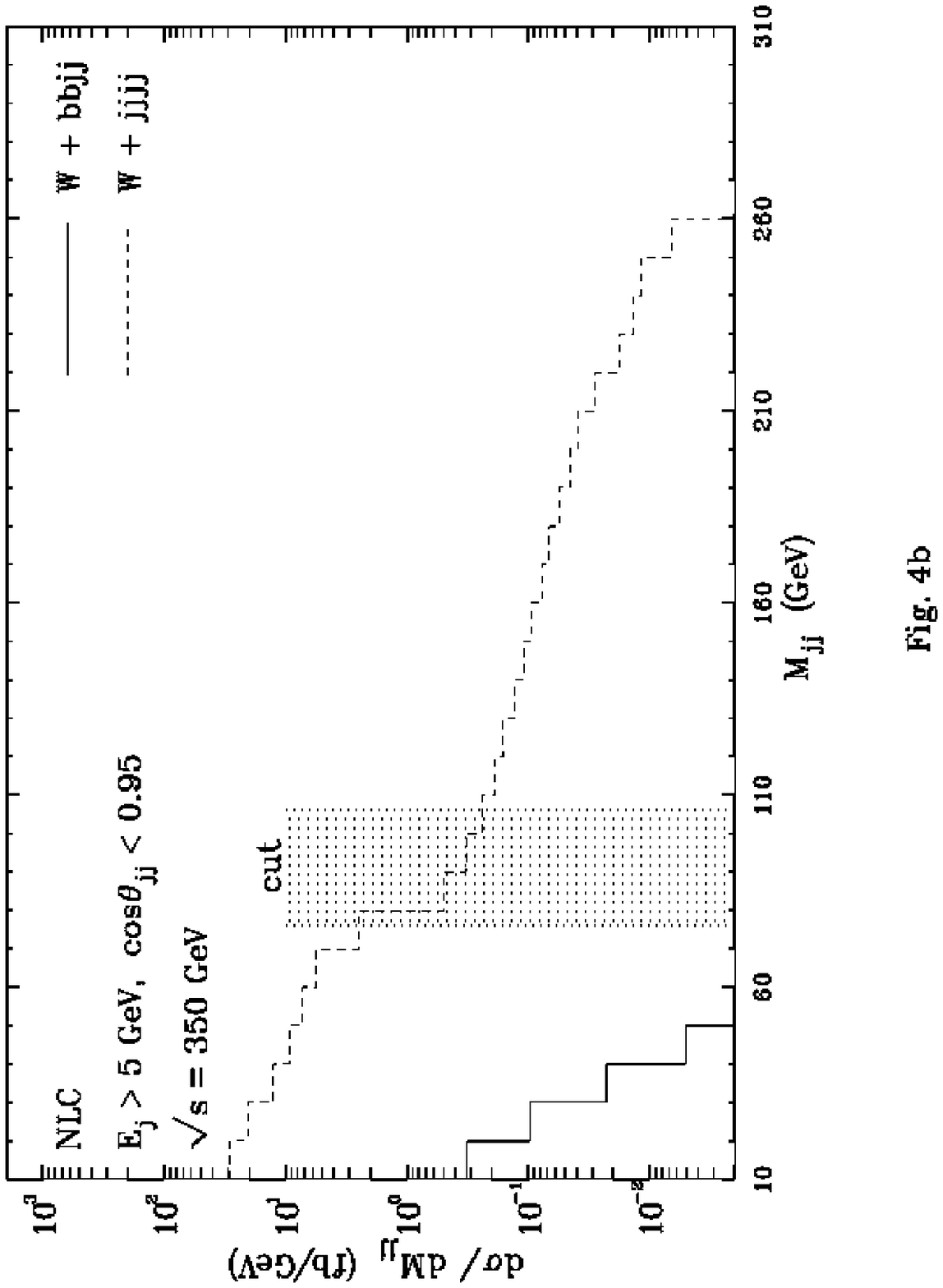,height=22cm}}  
\vspace*{2cm}
\end{figure}
\vfill
\clearpage
\begin{figure}[p]
~\epsfig{{file=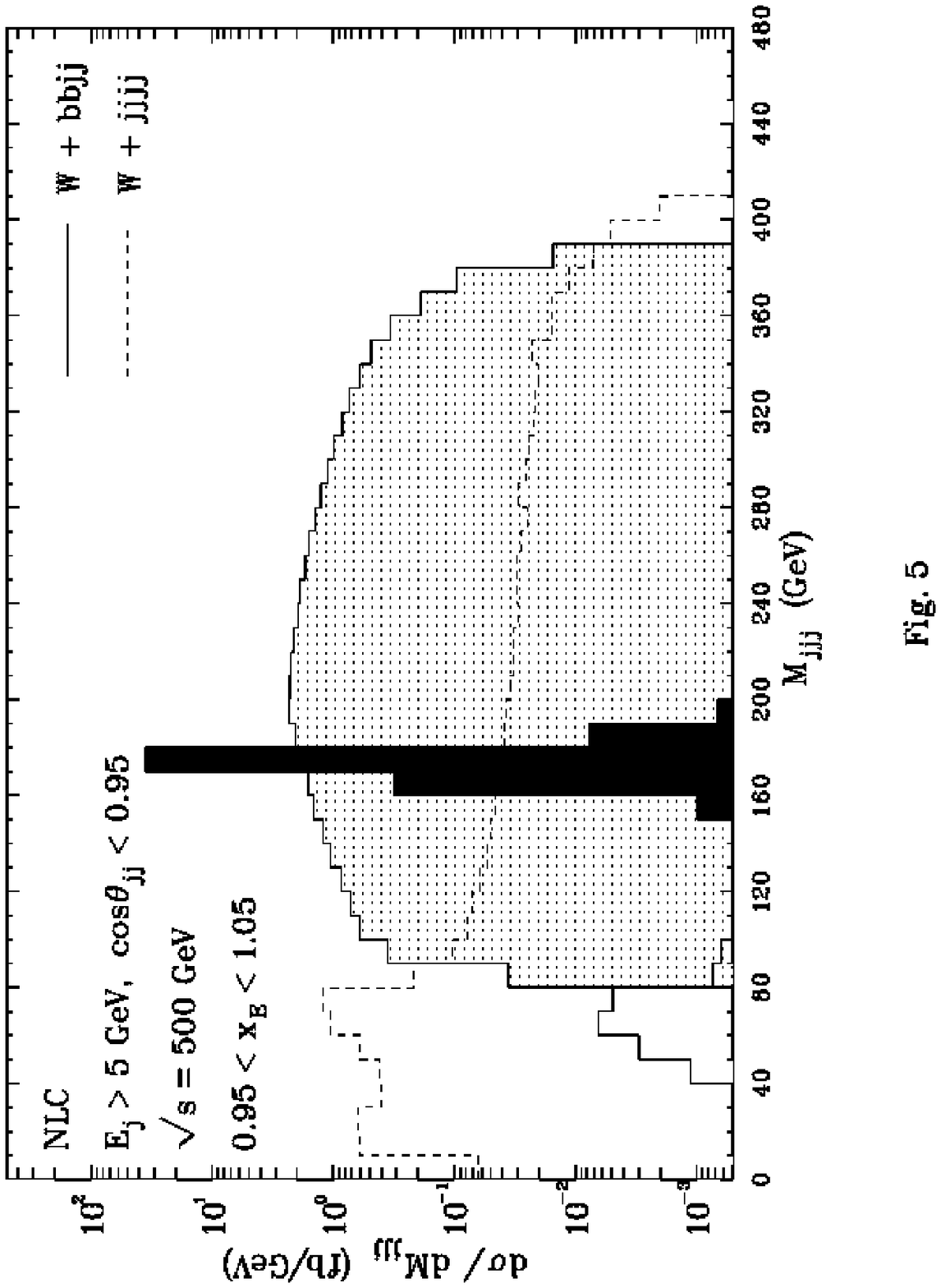,height=22cm}}  
\vspace*{2cm}
\end{figure}
\vfill
\clearpage
\begin{figure}[p]
~\epsfig{{file=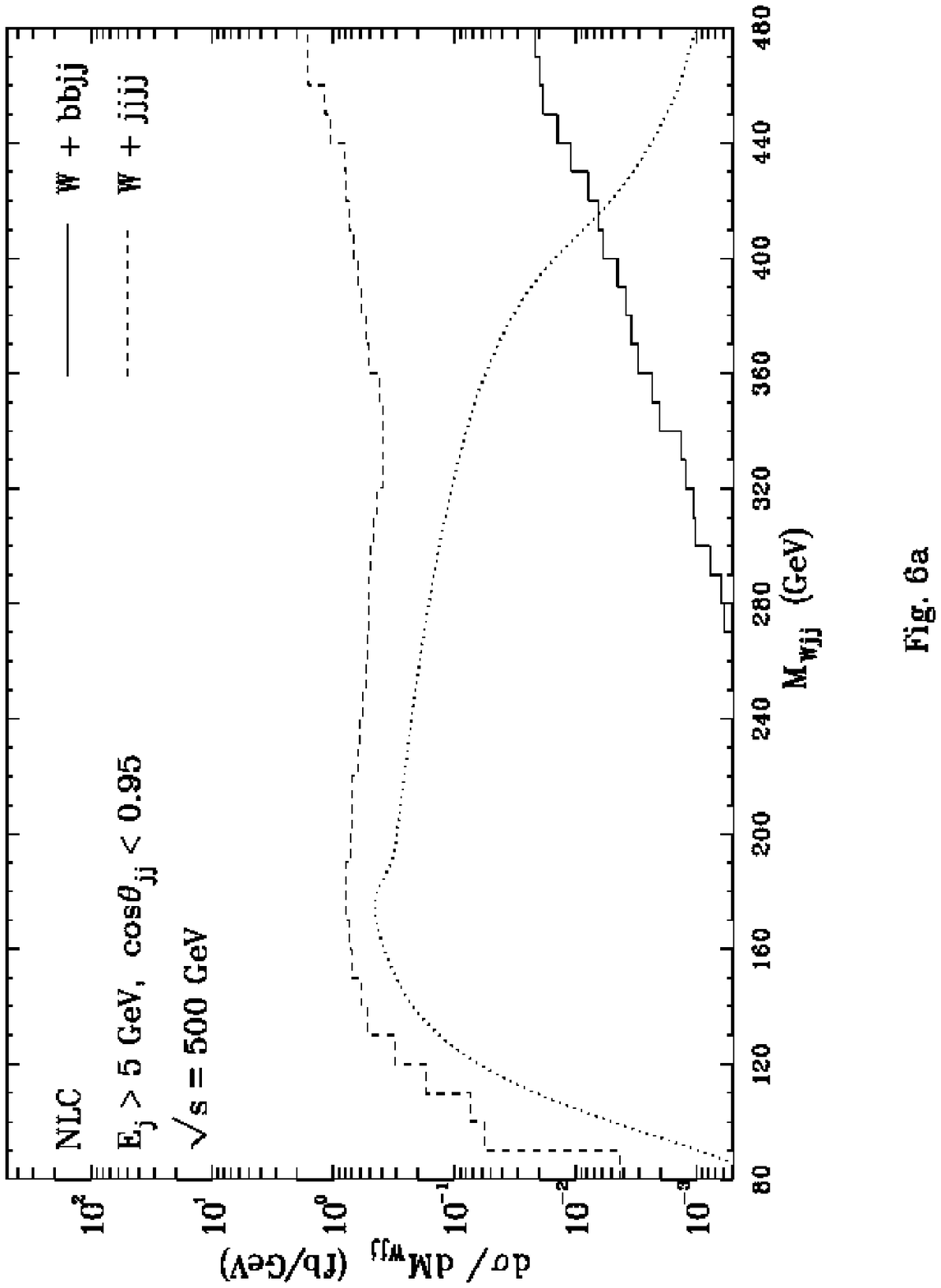,height=22cm}}  
\vspace*{2cm}
\end{figure}
\vfill
\clearpage
\begin{figure}[p]
~\epsfig{{file=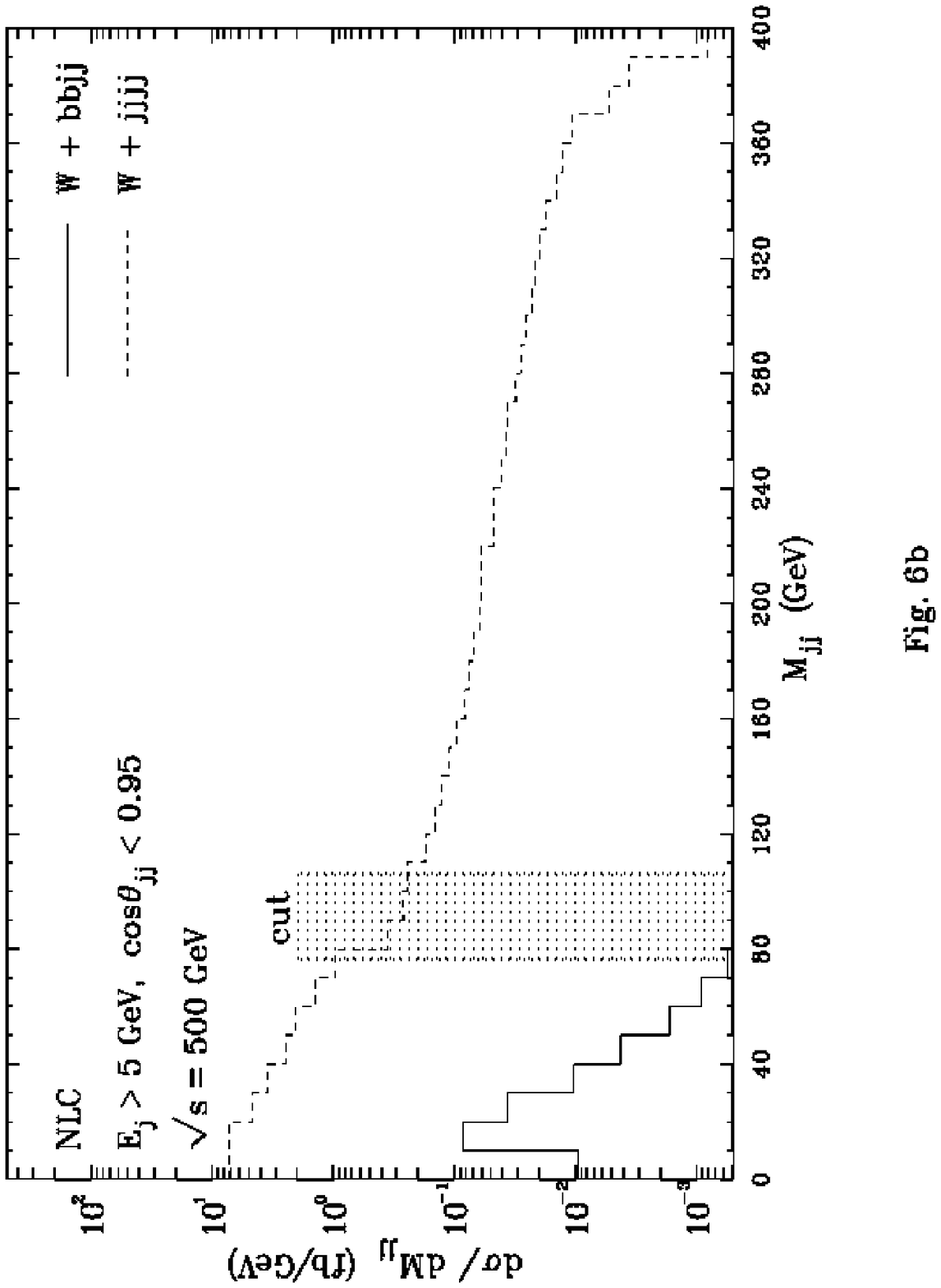,height=22cm}}  
\vspace*{2cm}
\end{figure}
\vfill
\clearpage

\end{document}